# Is phosphorene with intrinsic defect still an ideal anode material?

*Ruiqi Zhang [a], Xiaojun Wu [a, b, c], and Jinlong Yang [a, b, ***

[a] Hefei National Laboratory for Physical Sciences at the Microscale, University of Science and Technology of China, Hefei, Anhui 230026, China

[b] Synergetic Innovation Center of Quantum Information & Quantum Physics, University of Science and Technology of China, Hefei, Anhui 230026, China

[c] CAS Key Laboratory of Materials for Energy Conversion, Department of Materials Science and Engineering, University of Science and Technology of China, Hefei, Anhui 230026, China

**ABSTRACT** The diffusion of Li in electrode materials is key factor to charging/discharging rate capacity of Li-ion battery (LIB). Recently, two-dimensional phosphorene has been proposed as a very promising electrode material due to its ultrafast and directional lithium diffusion, as well as large energy capacity. Here, on the basis of density functional theory, we report that the intrinsic point defects, including vacancy and stone-wales defects, will block the directional ultrafast diffusion of lithium in phosphorene. On the defect-free phosphorene, diffusion of Li along the zigzag lattice direction is 1.6 billion times faster than along the armchair lattice direction, and 260 times faster than that in graphite. By introducing intrinsic vacancy and stone-wales defect, the diffusion energy barrier of Li along zigzag lattice direction increases sharply to the range of 0.17 ~ 0.49 eV, which block the ultrafast migration of lithium along the zigzag lattice direction. Meanwhile, the open circuit voltage increases with the emergence of defects, which is not suitable for anode materials. In addition, the formation energies of defects in phosphorene are much lower than those in graphene and silicene sheet, therefore, it is highly demanded to generate defect-free phosphorene for LIB applications.

**INTRODUCTION**

With the sharply increasing demands of power for consumer electronics, rechargeable lithium-ion battery (LIB) has attracted intensive attentions recently as an important energy storage device.[1–3] To obtain a high-performance LIB, it is highly desirable to develop novel electrode materials with long circle life, high energy, and charging/discharging rates capacity.[4–7] For the latter, the migration rate of lithium in electrode materials is one key, which is largely affected by the interaction between lithium and electrode materials.

In the past decades, graphite has been used as a conventional anode material, but its rates and energy capacity of lithium is still far away to meet the requirement of many applications, such as electric power vehicles. Alternatively, two-dimensional (2D) materials have been widely explored in recent years. For example, grapheme and graphene-like materials, including molybdenum disulfide ($MoS_2$) and vanadium disulfide ($VS_2$), have been considered as ideal electrode materials,[8–10] due to their high lithium storage capacity[11], novel electronic properties, and low energy barrier for lithium diffusion. For example, the theoretical maximum of specific capacity of graphene is ~774 mAh/g, and the energy barriers of lithium migration are as low as about 0.33, 0.25 and 0.22 eV in graphene, $MoS_2$ and $VS_2$, respectively, indicating their excellent performance in rate capacity.[12–14]

Recently, a new 2D material, phosphorene or few-layer black phosphorus (BP), has been isolated successfully by mechanical exfoliation from BP crystals, which exhibits a sizable direct band gap and ultrahigh carrier mobility.[15–17] Because of its distinctive structure, phosphorene presents a strongly anisotropic conducting behavior and optical conductivity in experiments,[18,19] which is different from other 2D material. It is well known that BP bulk is one of the most promising anode materials in LIBs due to its exceptionally high Li storage capacity (1279 mA h/g).[20] Shijun et al.[21] have demonstrated that phosphorene also have highly reversible capacity (~433 mA h/g), low open circuit voltage, small volume change and electrical conductivity of lithiated phosphorene, suitable for anode materials. Very recently, several recent theoretical works have reported that the diffusion of Li in phosphorene is ultrafast with an ignorable energy barrier of ~0.09 eV,[22,23] indicating that phosphorene is superior in ultrafast charging/discharging rate. In particular, the diffusion of Li in phosphorene is directional in one dimension, which behaviors significantly different from other 2D materials, such as graphene, $MoS_2$, and $VS_2$. These investigations indicate that phosphorene may be an ideal anode material for LIB with both high energy capacity and ultrafast charging/discharging rate.

However, it is well known that the diffusion of Li in 2D materials is affected by intrinsic defect, such as vacancy and stone-wales defects, partly due to the increased chemical reactivity in the defect regions.[13,24–26] In graphene, $MoS_2$ and $VS_2$ monolayer, the diffusion of Li is isotropic, where Li may "bypass" the defected region. For directional diffusion of Li in phosphorene, it raises an interesting question: Is phosphorene with intrinsic defects still an ideal anode material? To answer this question, a detailed study on the formation of intrinsic defect in phosphorene and diffusion of Li in phosphorene with intrinsic defects is urgent needed.

Here, we report our first-principles studies of the diffusion of Li atom in both perfect and defective phosphorene monolayer. Three kinds of intrinsic defects, including mono- and di-vacancy, as well as stone-wales defects, are considered. Our studies show that the diffusion of Li atom in perfect phosphorene monolayer is ultrafast and directional. Li atom prefers to diffuse along the zigzag direction with an ignorable energy barrier of 0.076 eV. In contrary, this value sharply increases to 0.490 eV on phosphorene with intrinsic defect, implying that the ultrafast directional diffusion of Li atom in phosphorene will be blocked. Moreover, our calculations show that the formation of energy of intrinsic defects is very low in phosphorene. Comparing with the defect-free phosphorene, the strong binding between Li and defected phosphorene increases its open circuit voltage. Therefore, compared with defect-free phosphorene, the phosphorene with defects is not a good choice for anode material with fast charging/discharging rate and new synthesis method should be developed to avoid the intrinsic defect for LIB applications.

**METHODS**

All calculations are performed with the density functional theory (DFT) method implemented in the VASP package.[27,28] The generalized gradient approximation of Perdew, Burke, and Ernzerhof (GGA-PBE) and projector augmented wave (PAW) potentials are used.[29] The kinetic energy cutoff are set to be 520 eV in the plane-wave expansion and ensures an accuracy of the energy of 1 meV/atom. A 5×5×1 supercell was adopted to analyze the Li diffusion and adsorption on defect phosphorene. Geometry structures are fully relaxed until energy and force are converged to $10^{-5}$ eV and 0.01 eV/Å, respectively. Dipole correction is employed to cancel the errors of electrostatic potential, atomic forces and total energy, caused by periodic boundary condition.[30] A large value ~15 Å of the vacuum region is used to avoid interaction between two adjacent periodic images. In order to estimate the charge transfer between Li and phosphorene, we adopt Bader charge analysis method, using the Bader program of Henkelman's group.[31,32] The climbing image nudged elastic band (CI-NEB) method[33,34] is used for minimum energy pathway (MEP) calculations of the penetration of Li atom diffusion on phosphorene. Four images are inserted between the initial and final states. All structures are fully relaxed until the convergence criteria of energy and force are less than $10^{-5}$ eV and 0.02 eV/Å, respectively.

**RESULTS AND DISCUSSION**

**Defects in Phosphorene**.

Unlike the flat structure of graphene, the phosphorene monolayer has a puckered honeycomb structure with each phosphorus atom forming three P-P covalently bonds with adjacent P atoms by sharing its *p* electrons, as shown in Fig.1(a). The optimized lattice constants are a = 3.30 Å, b = 4.62 Å, which is consistent with previous theoretical calculations.[17,19,35] Fig1.(b) shows the electronic band structure of perfect phosphorene, which is a direct-gap semiconductor with a band gap of 0.91 eV at Γ point in PBE calculation level, agreeing well with previous report.[17] These results will be compared with defected phosphorene, as shown later.

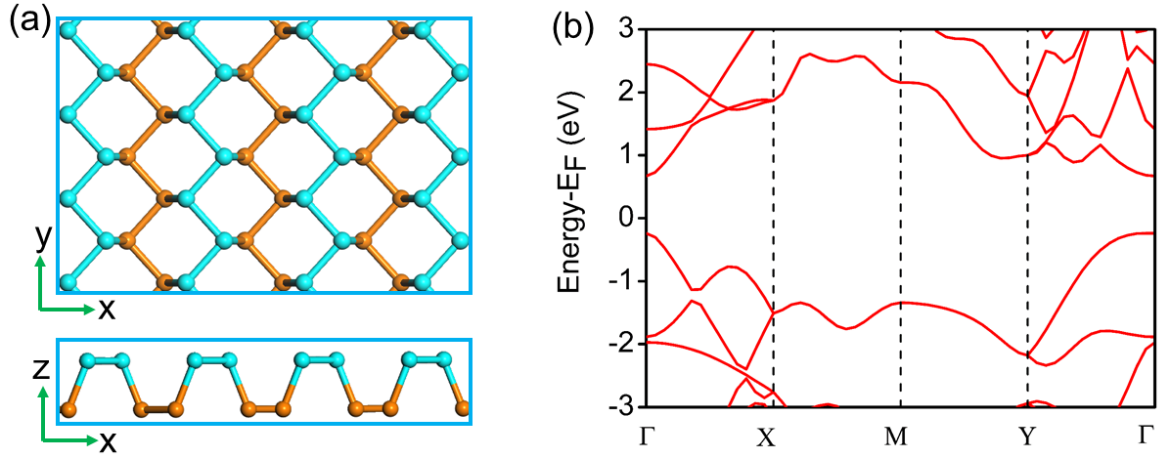

**Figure 1.** (a) The top view and side view of phosphorene. Atoms in different layers are depicted in cyan and orange. (b) Band structure of the pristine phosphorene calculated using PBE.

According to recent experimental work, phosphorene may not be as robust in air as other layered materials.[36,37] In fact, the structural defects may be produced in the process of synthesis. Several typical defects are considered, including Stone-Wales (SW), single and double vacancy (SV and DV) defects. The optimized structure of point defects in phosphorene are shown in Fig.2. As shown in Fig. 2 (a), four hexagons are transformed into two pentagons and two heptagons [SW (55-77) defect] by rotating one of the conplane P-P bonds by 90° and SW defect has a slightly higher formation energy. For the simplest defect in any material is missing lattice atom. SV defect in phosphorene is remove a P atom from the monolayer. Missing a P atom leads to the formation of a five-membered and a nine-membered ring [SV (5-9) defect]. The lowest-energy point defect is DV defect, which is composed of two buckled pentagons and one octagon [DV (585) defect]. Their band structure are depicted in Fig.2 (d), (e) and (f), respectively. According to our calculations, only SV defect introduces dangling bonds, which maybe the origin of p-type conductivity observed experimentally in phosphorene. Owing to their bonds are fully saturated in DV and SW defects, there no gap states appear.

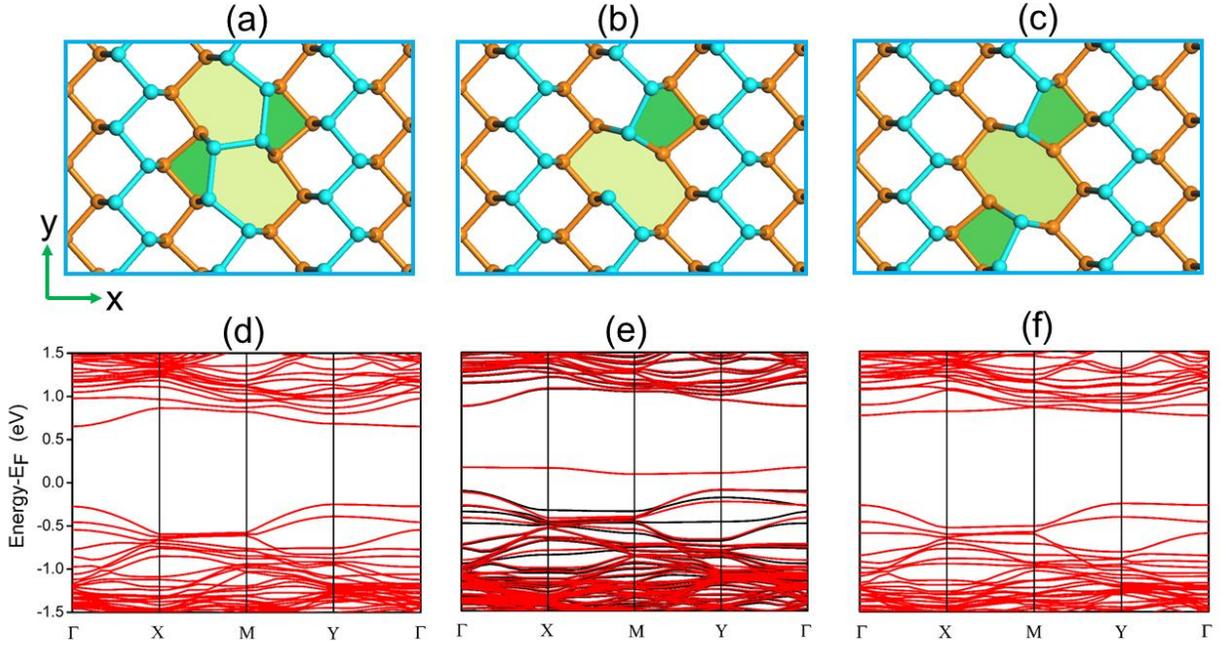

**Figure 2.** The top view of defect phosphorene and atoms in different layers are depicted in cyan and orange (a) SW defect, (b) SV defect, (c) DV defect. (d), (e), and (f) are their corresponding band structure calculated using PBE.

To assess the stability of defects in phosphorene, we calculate the formation energy $E_f$, which is defined as

$$E_f = E_{defect} - \frac{N-i}{N} E_{perfect} \quad (1)$$

, where $E_{defect}$ and $E_{perfect}$ are the total energy of the defected and defect-free phosphorene, respectively, $N$ is the number of atoms in the pristine unit cell, and $i$ is the number of atoms removed. As summarized in Table I, the calculated formation energy of SW, SV, and DV defects are 1.63, 1.29, and 1.28 eV per supercell, respectively. Among them, the SV and DV defects have smaller formation energy than that of SW defect, indicating that the former two are more easily generated in phosphorene. These values agrees well with previous report.[38,39] Moreoever, the formation energies of defects in phosphorene are smaller than those of same defects in graphene[40,41] and silicenen[42], as summarized in Table 1, indicating that these defects are quite easily formed in phosphorene. Therefore, it is crucial to investigate the interaction of lithium on the defective phosphorene for the application to LIBs.

Table 1. Formation energies $E_f$ (eV) of various defects in phosphorene, silicene and graphene

|  | SW | SV | DV |
| --- | --- | --- | --- |
| Phosphorene | 1.63 | 1.29 | 1.28 |
| Silicene | 2.09[a] | 3.01[a] | 3.70[a] |
| Graphene | 4.5-5.3[b] | 7.38-7.85[c] | 7.52-8.7[c] |

[a]Ref 42. [b]Ref 40. [c]Ref 40,41

**Lithium Adsorption on Phosphorene**

At first, we studied the adsorption of single Li atom on both perfect and defective phosphorene. The binding energy ($E_f$) of Li atom on phosphorene is defined as

$$E_f = (E_{nLi+P} - E_P - nE_{Li})/n \qquad (2)$$

, where $n$ represents the number of Li atoms in the system, $E_{Li}$ is the energy of metallic Li, $E_P$ is the total energy of the perfect or defective phosphorene, and $E_{nLi+P}$ refer to the total energy of system with *n* number of Li. Based on our definition, a more negative $E_f$ suggests a more stable system and the open circuit voltage (OCV) is related to the formation energy by,[43]

$$\text{OCV} \approx -E_f / e \qquad (3)$$

There are two stable adsorption sites for Li adsorption on the perfect phosphorene, the hollow site (H) above the center of the triangle consisting of three P atoms and bridge of P-P bond (B), as shown in Fig. 3(a). The binding energies are -0.23 and 0.30 eV for H and B sites, respectively, and the H site is the most stable adsorption site for Li atom. Besides these we also study the top site (T) directly above on a P atom. However, it is not a local minimum site for Li adsorption and Li atom on the T site will move to H site after a full structural relaxation.

On the defected phosphorene, Li atom prefers to adsorb in the hollow site, as depicted in Fig.3. Four parameters are defined to evaluate the adsorption behaviors of Li atom on different defects. The first parameter is the perpendicular distance between Li atom and phosphorene (*Z*). The second one is the nearest distance between Li atom and phosphorus (*R*). The third one is the *OCV* and the last one is the charge transfer parameter (*Q*) calculated by Bader charge analysis method. All the results are list in Table 2.

From the calculated OCV, it is obviously that the defects increased the value of OCV. However, for anode materials, a suitable OCV with low value is important for LIB applications.[44] The value of *Z* for the DV defect is only 0.16 Å, indicating that the lithium atom is trapped inside the vacancy, as shown in Fig.3 (d). There a significant charge transfer between Li atom and phosphorene, and the value of Q doesn't vary much with the calculated binding energy.

**Table 2.** Perpendicular height $Z$, the nearest distance between lithium and phosphorus $R$, binding energy $E_b$, and charge transfer parameter $Q$ are summarized for the perfect and defective phosphorene.

|         | Pristine | SW   | SV   | DV   |
|---------|----------|------|------|------|
| $Z$ (Å) | 1.51     | 1.33 | 1.24 | 0.16 |
| $R$ (Å) | 2.46     | 2.52 | 2.39 | 2.58 |
| $OCV$ (V) | 0.23   | 0.35 | 1.36 | 0.76 |
| $Q$ (e) | 0.59     | 0.38 | 0.50 | 0.29 |

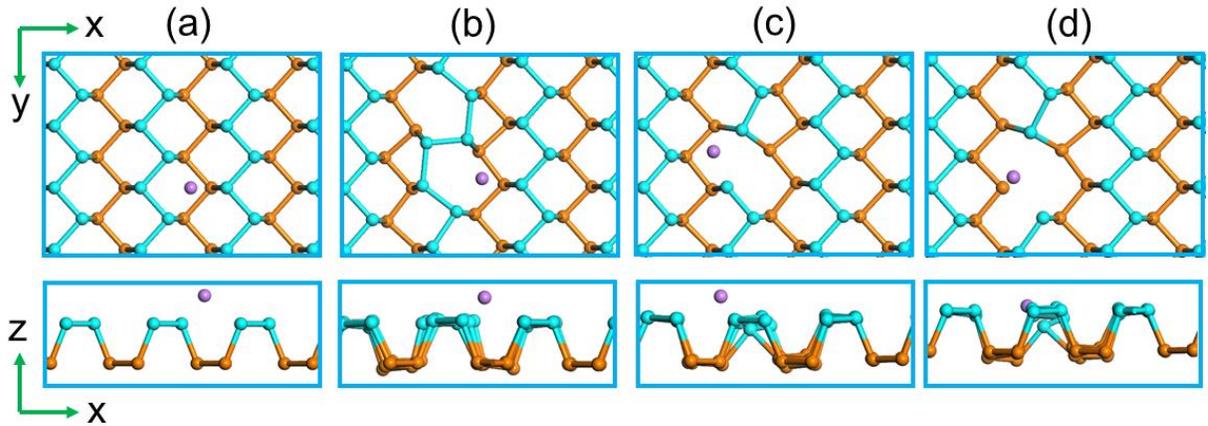

**Figure 3**. Top and side views of lithium adsorbed on (a) pristine, (b) SW, (c) SV, and (d) DV phosphorene. Phosphorus atoms in different layers are depicted in cyan and orange and lithium atom is depicted in purple.

**Diffusion of Li atom**

It is well known that the performance of anode materials largely depends on the diffusion rate of Li atom. Next, we explored the in-plane diffusion of Li atom on both perfect and defected phosphorenes. As mentioned above, the most stable adsorption site of Li atom is H site for the perfect phosphorene. Three pathways for Li diffusion between two H sites are considered, as shown in Figure 4 ($a_1$). The first path is migration from $H_0$ to $H_1$ site ($H_0 \rightarrow H_1$) over a T site (Path A). The calculated migration energy barrier is 0.624 eV. The second one is migrating from $H_0$ to $H_2$ over a B site ($H_0 \rightarrow H_2$) with an energy barrier of 0.590 eV (Path B). The third one is migrating along zigzag direction ($H_0 \rightarrow H_3$) with an ignorable energy barrier of 0.076 eV (Path C). Clearly, the diffusion of Li atom on the perfect phosphorene is anisotropic, which is consistent with previous report[22,23] This anisotropic diffusion behavior originates from the anisotropic structure of the perfect phosphorene. Moreover, the energy barrier for the diffusion of Li on the perfect phosphorene is much lower than that of other 2D materials. For example, the diffusion energy barriers of Li atom are 0.25 eV on $MoS_2$,[13,14] 0.22 eV on $VS_2$,[14] 0.33 eV on graphene.[12]

According to the Arrhenius equation, the diffusion constant (D) can be expressed by

$$D \sim \exp\left(\frac{-E_a}{k_B T}\right) \qquad (3)$$

, where $E_a$ and $k_B$ are the activation energy barrier and Boltzmann constant, respectively. Taking the pristine phosphorene for example, it is estimated that the Li mobility along Path C is about $1.6 \times 10^9$ times faster than along Path A at room temperature and more than 260 times faster than in graphite (the energy barrier for Li diffusion in graphite is 0.22eV).[14] Therefore, the diffusion of Li atom on the perfect phosphorene is directional and ultrafast with ignorable diffusion energy barrier.

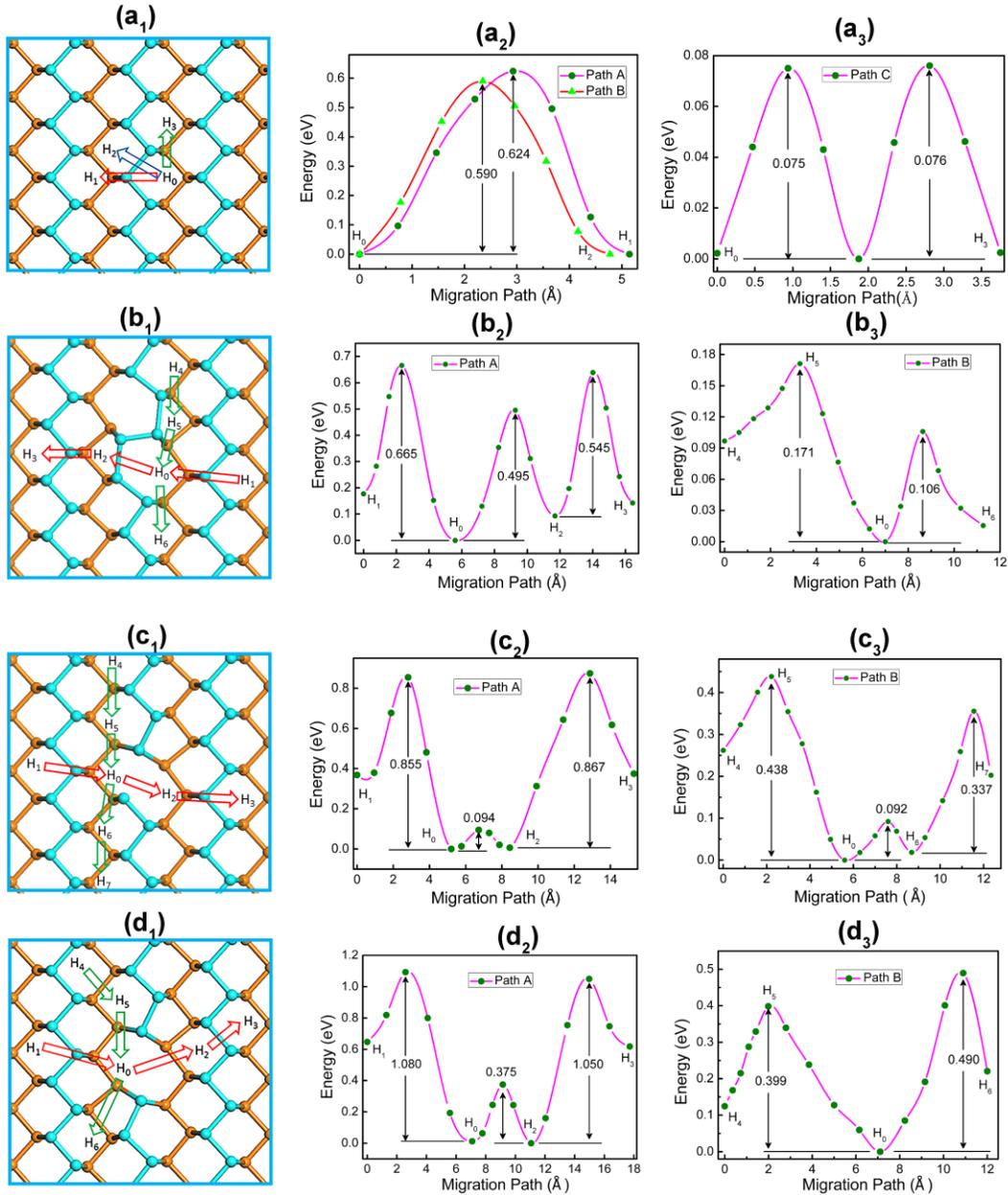

**Figure 4**. Schematic representations and potential-energy curves of Li diffusion on pristine and defective phosphorene ($a_1, a_2, a_3$) Li diffusion on pristine phosphorene following three paths (Path A: $H_0 \rightarrow H_1$, Path B: $H_0 \rightarrow H_3$, and Path C: $H_0 \rightarrow H_3$); ($b_1, b_2, b_3$) Li diffusion on SW phosphorene

along two paths (Path A: $H_1 \rightarrow H_0 \rightarrow H_2 \rightarrow H_3$ and Path B: $H_4 \rightarrow H_5 \rightarrow H_0 \rightarrow H_6$); ($c_1, c_2, c_3$) Li diffusion on SV phosphorene along two paths (Path A: $H_1 \rightarrow H_0 \rightarrow H_2 \rightarrow H_3$ and Path B: $H_4 \rightarrow H_5 \rightarrow H_0 \rightarrow H_6$) and ($d_1, d_2, d_3$) Li diffusion on DV phosphorene along two paths (Path A: $H_1 \rightarrow H_0 \rightarrow H_2 \rightarrow H_3$ and Path B: $H_4 \rightarrow H_5 \rightarrow H_0 \rightarrow H_6$).

Next, the diffusion of Li atom on the defected phosphorene is investigated. As shown in Fig. 4, two pathways around the defect are considered, including the migration along armchair and zigzag directions. The whole migration pathways including trapping in the defects and escaping from the defect regions are considered. From the calculated minimum energy pathway of Li's diffusion, it can be found that the intrinsic point defects create a potential trap in the defected region, which increases the diffusion energy barrier of Li atom.

For SW defect, two pathways are studied, including Path A: $H_1 \rightarrow H_0 \rightarrow H_2 \rightarrow H_3$ and Path B: $H_4 \rightarrow H_5 \rightarrow H_0 \rightarrow H_6$. Obviously, the Li diffusion along Path A has a relatively higher energy barrier than that along Path B. The highest energy difference along Path A is from $H_1$ to $H_2$ with a value of 0.665 eV. Along Path B, the energy barrier is about 0.171 eV. The calculated minimum energy pathway (MEP) are depicted in Figure 4($b_2$) and ($b_3$). For SV defect, we also considered two paths for the diffusion of Li atom, *i.e.* Path A: $H_1 \rightarrow H_0 \rightarrow H_2 \rightarrow H_3$ and Path B: $H_4 \rightarrow H_5 \rightarrow H_0 \rightarrow H_6$, as shown in Fig. 4 ($c_1$). As shown in Fig. 4($c_2$) and ($c_3$), 0.855 and 0.867 eV are respectively needed to overcome the energy barriers for the diffusion of Li atom from $H_1$ to $H_0$ and $H_2$ to $H_3$ in Path A, which are higher than that along Path B (0.438 and 0.337 eV). Note that although the energy barrier for the diffusion of Li atom in the defect region is only ~0.09 eV ($H_0 \rightarrow H_2$), the energy barrier for Li atom escaping from the defected region is still very high. For the DV defect, the energy barrier increases to 1.080 eV for the diffusion of Li atom between $H_1$ and $H_0$ sites along the Path A. Meanwhile, the energy barrier increases to 0.490 eV for the diffusion between $H_0$ and $H_6$ sites along the Path B. These results indicate that the diffusion of Li atom in the defected phosphorene is still anisotropic with a significantly increased energy barrier.

According to the equation (3), for the defected phosphorene, the increased energy barrier definitely reduces the diffusion rate of Li atom in phosphorene. For SW defect, the energy barrier along the zigzag direction is 0.171 eV, which is still smaller than those of other 2D materials. However, the energy barrier of the phosphorene with vacancy defect, which has the lowest formation energy, is much larger than that of other 2D materials. As the diffusion of Li atom on phosphorene is directional, the existence of intrinsic point defect will block the diffusion of Li atom. This is quite different from the diffusion on graphene and other 2D materials, where the diffusion of Li atom is isotropic. Since the formation energy of vacancy defect is lower than that of SW defects. Therefore, the defected phosphorene with vacancy defects will lose the advantage of rapid discharge and charge and it is necessary to avoid the formation of defects in phosphorene for LIBs applications.

**CONCLUSION**

In conclusion, on the basis of DFT calculations, we performed a theoretical study on the adsorption and diffusion behavior of Li atom on both defect-free and defective phosphorene. The calculated formation energies of SW, SV, and DV are 1.63, 1.29, and 1.28 eV, indicating the intrinsic point defects are easier formed in phosphorene than graphene and silicene. At the same time, the OCV increased greatly as defects appearing. On the defect-free phosphorene monolayer, Li atom preform to adsorb in the hollow site. The diffusion of Li atom is directional, where the diffusion constant along the zigzag direction is 1.6 billion times faster than that along the armchair directions, and more than 260 times faster than that in graphite. However, on the defective phosphorene, the diffusion energy barrier of Li atom along the zigzag direction increase to 0.171, 0.438, and 0.490 eV for SW, SV, and DV defects, which will block the directional diffusion of Li atom in phosphorene. Therefore, it is essential to fabricate defect-free phosphorene in experiment for its application LIBs with high energy capacity and ultrafast charging/discharging rates.

Very recently, we noticed that Gencai *et al.*[44] have investigated the application of defected phosphorene in anode material of LIB. They show that phosphorene with vacancy defect increases the adsorption energy of Li atom, and the diffusion of Li atom in the defected region is only 0.13 eV, which consistent with our calculations (0.09 eV in the defected region). However, they don't consider the migration of Li atom from the defect region to perfect region, which the energy barrier increases to at least 0.337 and 0.855 eV for Path B and A (Figure 4($c_2$) and ($c_3$)), respectively. Meanwhile, a small supercell (4×4×1) used in previous work will affect the calculation results.

**AUTHOR INFORMATION**

**Corresponding Author**

* E-mail: jlyang@ustc.edu.cn. Phone: +86-551-63606408. Fax: +86-551-63603748 (J. Y.).

**Author Contributions**

The manuscript was written through contributions of all authors. All authors have given approval to the final version of the manuscript.

**ACKNOWLEDGMENT**

This work is partially supported by the National Key Basic Research Program (2011CB921404, 2012CB922001), by NSFC (21121003, 91021004, 21233007, 21203099, 21273210, 21421063, 51172223), by Strategic Priority Research Program of CAS (XDB01020300), the Fundamental Research Funds for the Central Universities (WK2060190025, WK2060140014), and by USTCSCC, SCCAS, Tianjin, and Shanghai Supercomputer Centers. The authors would like to thank Dr. Zhiwen Zhuo from the Department of Materials Science and Engineering, USTC in Hefei, China, for valuable discussions.